\documentclass[12pt]{iopart}
\usepackage[dvips]{graphicx}
\usepackage{iopams,setstack}

\begin{document}
\article[$\phi^4$ Theory with Non-Linear Curvature Coupling]{}
{Radiative Symmetry Breaking and Dynamical Origin of Cosmological 
Constant in $\phi^4$ Theory with Non-Linear Curvature Coupling}
\author{
T Inagaki
}
\address{
Information Media Center, Hiroshima University,\\
1-7-1 Kagamiyama, Higashi-Hiroshima, Hiroshima 739-8521, JAPAN}
\eads{\mailto{inagaki@hiroshima-u.ac.jp}}

\begin{abstract}
A scalar self-interacting theory non-linearly coupled with some power 
of the curvature have a possibility to explain the current smallness 
of the cosmological constant. In Ref.\cite{INO} a symmetry property 
of the model has been studied in an arbitrary dimensions and a solution 
of the cosmological constant problem has been found in four-dimensions. 
Here one concentrate on a massless scalar field in the four-dimensional 
Fridmann-Robertson-Walker (FRW) spacetime with flat spatial part. One 
apply the same analysis as Ref.\cite{INO} and show the phase structure 
of radiative symmetry breaking. One also review a dynamical resolution 
of the cosmological constant problem.
\end{abstract}


\maketitle

\section{Introduction}
One of the most important and mysterious problem in the present cosmology 
is the origin of the dark energy. It has an anti-gravity contribution to 
accelerate the universe expansion. Even in particle physics much interest 
has been paid to the problem and new field theoretical models are considered 
to explain the acceleration of our Universe. A simple candidate of the dark 
energy is found in the vacuum energy or cosmological constant. However, the 
scale of the dark energy is about 120 orders of magnitude smaller than the 
Planck scale. We can not avoid fine tuning to obtain a suitable scale vacuum 
energy in most of particle physics models. 

A interesting possibility to solve the dark energy problem is found in a 
modification of the gravity-matter coupling. S. Nojiri and S. D. Odintsov 
have proposed the class of dark energy models where dark energy field couples 
with some power of the curvature \cite{NO} (see also related model in 
\cite{ANO}). The model naturally resolves the problem of dark energy 
dominance in the current universe. In Ref.\cite{INO} the symmetry 
properties of such models have been studied on the example of scalar 
self-interacting theory with non-linear curvature coupling in arbitrary 
dimensions. It is found that the phase structure of the model strongly 
depends on the curvature power in $3.8$ spacetime dimensions. 

In the present paper we focus on the four-dimensional limit of the theory 
and continue the study of scalar self-interacting theory with non-linear 
curvature coupling. It should be noted that the theory is not standard one, 
in the sense that it is not multiplicatively renormalizable in curved 
spacetime \cite{BOS}. We regards it as an effective theory steaming from 
a more fundamental theory at Planck scale. In our analysis we neglect the 
quantum gravity effects. We also assume that the spacetime curved slowly 
and neglect the higher order terms about curvature. It is appropriate for 
the study of quantum effects in the inflationary universe and also in 
late-time, dark energy universe. 

First we introduce the scalar theory non-linearly coupled with some power 
of the curvature. To find the properties at the weak curvature limit we 
apply the Riemann normal coordinate expansion and evaluate the effective 
Lagrangian. The one-loop effective Lagrangian is obtained in close analogy 
with multiplicatively renormalizable theories. Next we consider the massless 
scalar field and investigate radiative symmetry breaking which is caused 
by the loop corrections of the scalar field. In our theory the 
expectation value of the scalar field corresponds to an order parameter 
of radiative symmetry breaking. We evaluate the effective Lagrangian
numerically in four dimensions and find the phase structure of the 
theory in four dimensions. Finally we show the solution of the Einstein 
equation and discuss dynamical resolution of the cosmological constant 
problem. 

\section{$\phi^4$ theory with non-linear curvature coupling}

As is well-known, $\phi^{4}$ theory is one of the simplest models where 
the spontaneous symmetry breaking takes place. It is more instructive to 
consider the $\phi^{4}$ theory as a prototype model to study an influence
of a non-linear curvature coupling.
Here we extend the $\phi^{4}$ theory to include a coupling with some power 
of curvature:
\begin{equation}
    S = \int d^D x \ \sqrt{-g} \left[\frac{1}{2\kappa^2}R
        + \left(\frac{R}{M^2}\right)^\alpha {\cal L}_d \right] .
\label{a0}
\end{equation}
where $M$ is an arbitrary mass scale, $g$ is the determinant of the metric
tensor $g_{\mu\nu}$ and ${\cal L}_d$ is the ordinary Lagrangian density of
the $\phi^{4}$ theory,
\begin{equation}
    {\cal L}_d(\phi_0) = \frac{1}{2}\phi_{0;\mu}{\phi_0}^{;\mu}
    - \frac{\xi_0 R}{2} \phi_0^2
    +\frac{\mu_0^2}{2}\phi_0^2
    -\frac{\lambda_0}{4!}\phi_0^4,
\label{I2}
\end{equation}
where $\phi_0$ is a real scalar field. Our 
sign convention for the metric is $(+---\cdots)$.

The action (\ref{a0}) is invariant under the discrete transformation,
$\phi \rightarrow -\phi$.
A non-vanishing expectation value for the field $\phi$ breaks
this discrete $Z_2$ symmetry spontaneously. The expectation value for 
$\phi$ is found by observing the stationary point of the effective 
action $\Gamma[\phi_c]$. 

In a constant curvature spacetime the non-linear curvature coupling
disappears by the transformation
\begin{eqnarray}
     \phi \rightarrow \left(\frac{R}{M^2}\right)^{-\alpha/2} \phi,
\ \ \ \ \ 
     \lambda_0 \rightarrow \left(\frac{R}{M^2}\right)^{\alpha} \lambda_0 .
\end{eqnarray}
Hence a non-trivial effect of the non-linear curvature coupling will be 
found only in a non-static and/or inhomogeneous spacetime.

Here we adopt the Riemann normal coordinate expansion \cite{P, BP, PT}
and evaluate the one-loop effective Lagrangian.
It is divergent in four dimensions. To obtain the finite result we have to 
renormalize the theory. Here we impose the following renormalization conditions
\begin{equation}
    \left. \frac{\partial^2 \Gamma[\phi]}{\partial \phi^2} \right|_{\phi=0}
    \equiv \left(\frac{R}{M^2}\right)^\alpha (\mu_r^2 - \xi_r R) ,
\ \ \ 
    \left. \frac{\partial^4 \Gamma[\phi]}{\partial \phi^4} \right|_{\phi=M}
    \equiv -\left(\frac{R}{M^2}\right)^\alpha \lambda_r ,
\label{def:lr}
\end{equation}
where $M$ is the renormalization scales.
From these conditions one obtains the renormalized 
parameters $\mu_r$, $\xi_r$ and $\lambda_r$.
By using these renormalized parameters the effective Lagrangian density
reads at the four-dimensional limit, 
\begin{eqnarray}
    {\cal L}_{eff}^{4D} &=&
        \frac{1}{2\kappa^2}R
        + \left(\frac{R}{M^2}\right)^\alpha \left(
        \frac{1}{2}\phi_{;\mu}{\phi}^{;\mu}
        - \frac{\xi_r R}{2} \phi^2
        +\frac{\mu_r^2}{2}\phi^2
        -\frac{\lambda_r}{4!}\phi^4\right)
\nonumber \\
    &&  +\frac{\hbar}{128\pi^2}\left[
        \lambda_r^2 \phi^4 \left(
        \frac{\lambda_r M^2}{\chi^2(M^2)}
        -\frac{1}{6}\frac{\lambda_r^2 M^4}{\chi^2(M^2)}
        \right)
        +\lambda_r\phi^2 \chi^2(0)
        \right.
\nonumber \\
    &&  +\frac{3}{4}\lambda_r^2\phi^4
        -2(\chi^2(\phi))^2\ln\frac{\chi^2(\phi^2)}{\chi^2(0)}
        -\frac{1}{2}\lambda_r^2\phi^4 \ln \frac{\chi^2(0)}{\chi^2(M^2)}
\label{alrd4} \\
    &&  -\left\{
        \frac{\lambda_r^2}{4}\phi^4 \left(
        \frac{1}{\chi^2(M^2)}
        -\frac{2\lambda_r M^2}{(\chi^2(M^2))^2}
        +\frac{2}{3}\frac{\lambda_r^2 M^4}{(\chi^2(M^2))^3}
        \right)\right.
\nonumber \\
    &&  \left.\left. +\lambda_r\phi^2
        -2\chi^2(\phi^2)\ln\frac{\chi^2(\phi^2)}{\chi^2(0)}\right\}
        \mbox{tr}\xi \right] .
\nonumber
\end{eqnarray}
with
\begin{eqnarray}
    \mbox{tr} \xi &=& \alpha \left( \frac{\square R}{R}
                   -\frac{R^{;\mu}R_{;\mu}}{R^2} \right) ,
\\
    \chi^2(\phi^2) & = & - \mu_0^2 +\frac{\lambda_0}{2}\phi^2
             + \left(\xi_0-\frac{1}{6}\right) R
             -\frac{\alpha}{2}\left(\frac{\alpha}{2}-1\right)
              \frac{R^{;\mu}R_{;\mu}}{R^2}
             -\frac{\alpha}{2}\frac{\square R}{R} ,
\end{eqnarray}
Therefore the ultra-violet divergences disappear after the one-loop 
renormalization. 
Of course, the question of dependence from the regularization in such 
effective models remains at higher order.

\section{Radiative symmetry breaking in four-dimensional FRW spacetime}

It is expected that the non-linear curvature coupling in our model leads to
non-trivial consequences for the phase structure in a non-static spacetime.
It is more interesting to consider the radiative symmetry breaking at 
$\mu_r=0$. In this case spontaneous symmetry breaking can not take place 
on the classical level of the theory. It is expected that the radiative 
correction plays an essential role and the curvature effect is more 
important for symmetry properties. 

Here we consider the model in the spatially flat FRW spacetime in four 
dimensions. It is defined by the metric
\begin{equation}
    ds^2 = dt^2 - a^2(t)
    \left[dr^2+r^2 d \Omega_{2}
    \right] .
\end{equation}
The time dependence of the scale factor is assumed to be $a(t)=a_0 t^{h_0}$. 
All mass scales are normalized by an arbitrary mass scale $M$ and
$\hbar =1$. 

First we consider the stationary and spatially homogeneous $\phi$.
In this case the kinetic term of $\phi$ disappears. One can define the 
effective potential by the opposite sign of the effective Lagrangian,
$V(\phi)\equiv -{\cal L}^{4D}_{eff}$.
The expectation value of the field $\phi$ is determined by observing the 
minimum of the effective potential. The effective potential develops a 
small imaginary part for a negative $\chi(\phi)$. Below we evaluate a real 
part of the effective potential to find the ground state.

\begin{figure}[t]
\begin{center}
\begin{minipage}{6.8cm}
\begin{center}
\includegraphics[width=\linewidth]{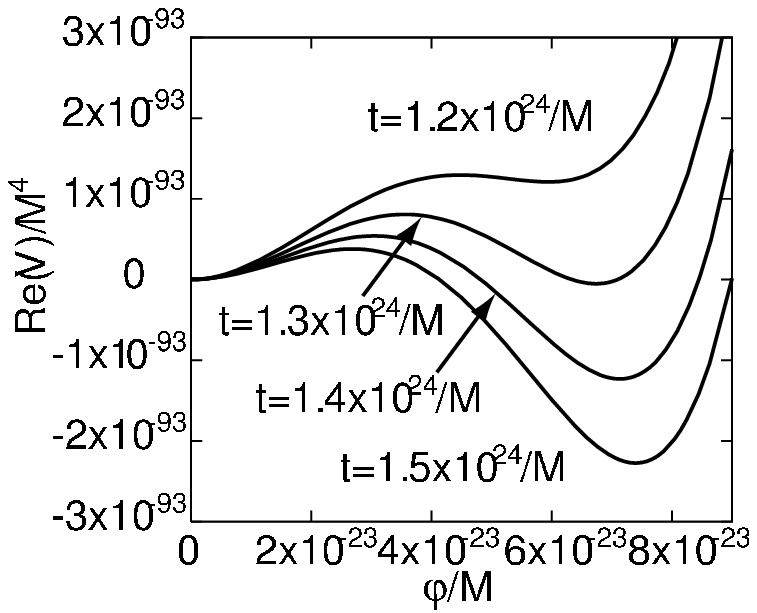}
(a) $\alpha$=0, $\xi=\xi_{\mbox{\scriptsize conformal}}$
\end{center}
\end{minipage}
\begin{minipage}{6.8cm}
\begin{center}
\includegraphics[width=\linewidth]{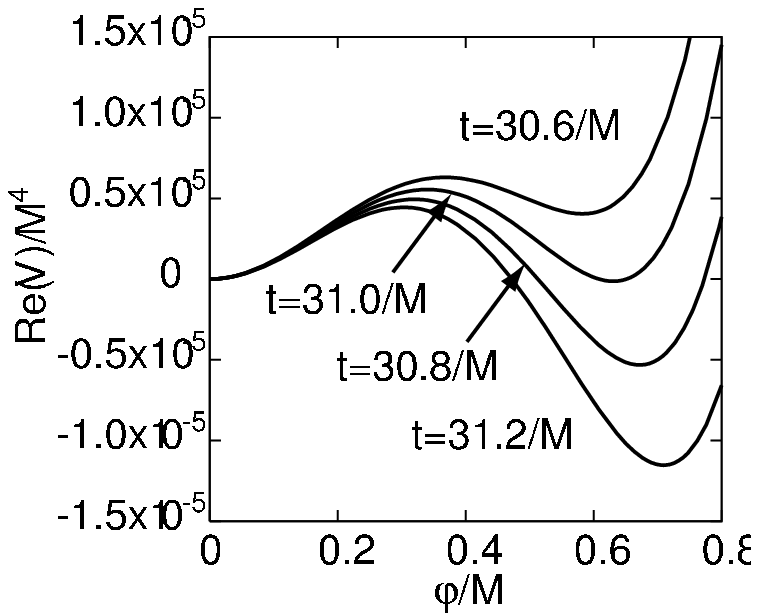}
(b) $\alpha$=1, $\xi=\xi_{\mbox{\scriptsize conformal}}$
\end{center}
\end{minipage}
\begin{minipage}{6.8cm}
\begin{center}
\includegraphics[width=\linewidth]{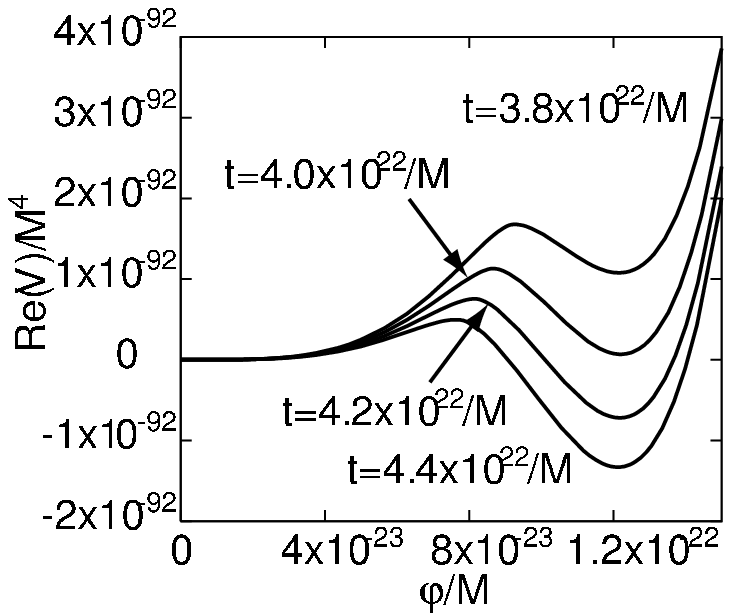}
(c) $\alpha$=0, $\xi=0$
\end{center}
\end{minipage}
\begin{minipage}{6.8cm}
\begin{center}
\includegraphics[width=\linewidth]{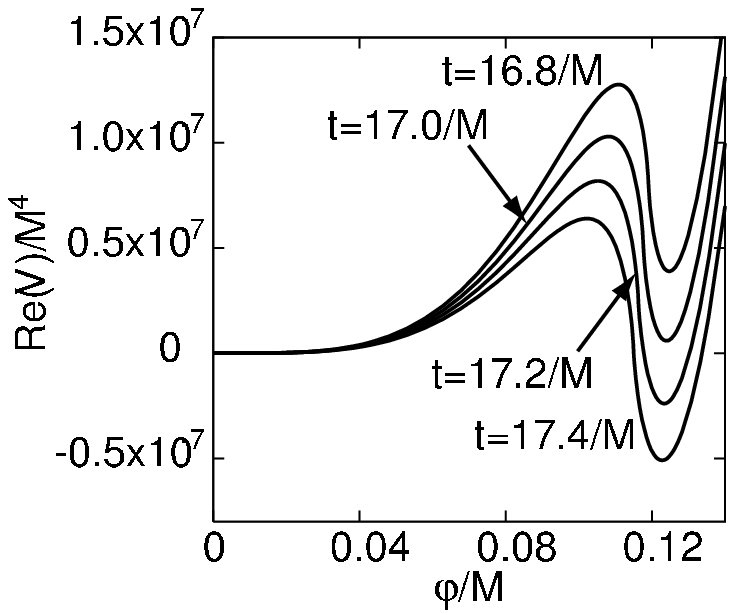}
(d) $\alpha$=1, $\xi=0$
\end{center}
\end{minipage}
\caption{\label{minimu0c2}Behaviour of the effective potential
for $h_0=2$, $\mu_r=0$, $\lambda=1$ and $D=4$.}
\end{center}
\end{figure}

\begin{figure}[t]
\begin{center}
\begin{minipage}{6.8cm}
\begin{center}
\includegraphics[width=\linewidth]{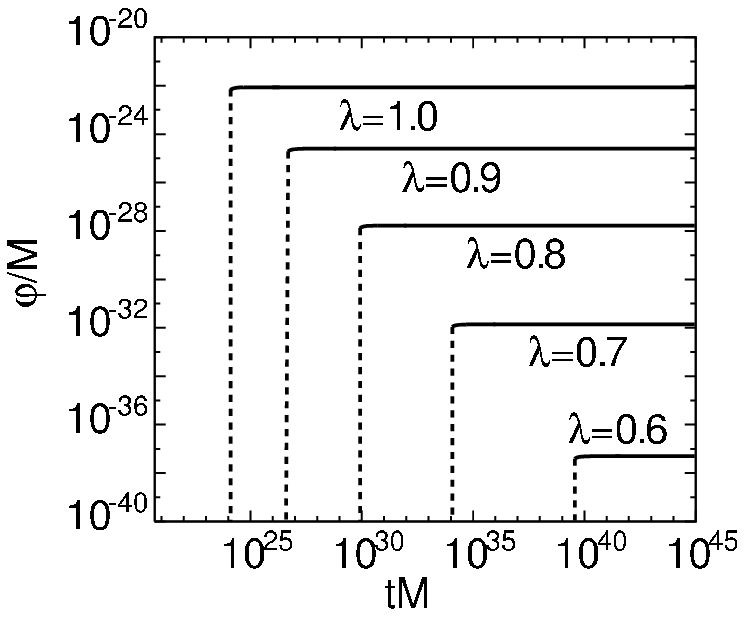}
(a) $\alpha$=0, $\xi=\xi_{\mbox{\scriptsize conformal}}$
\end{center}
\end{minipage}
\begin{minipage}{6.8cm}
\begin{center}
\includegraphics[width=\linewidth]{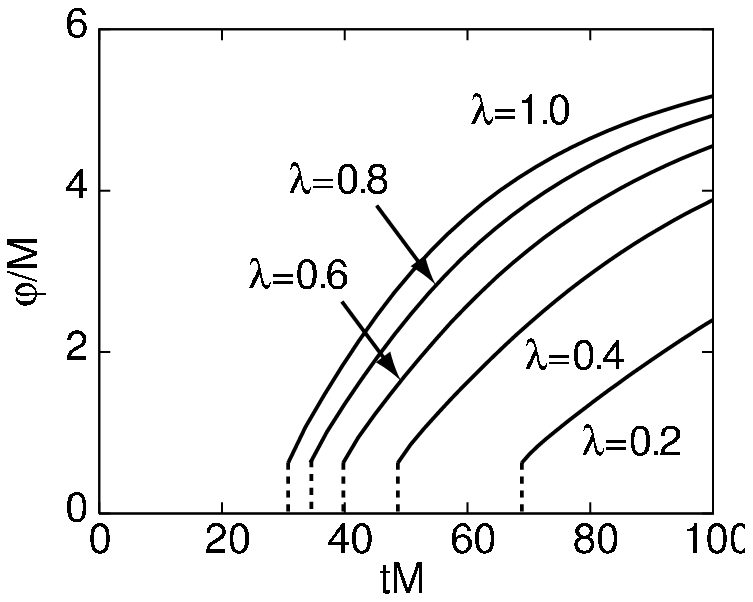}
(b) $\alpha$=1, $\xi=\xi_{\mbox{\scriptsize conformal}}$
\end{center}
\end{minipage}
\begin{minipage}{6.8cm}
\begin{center}
\includegraphics[width=\linewidth]{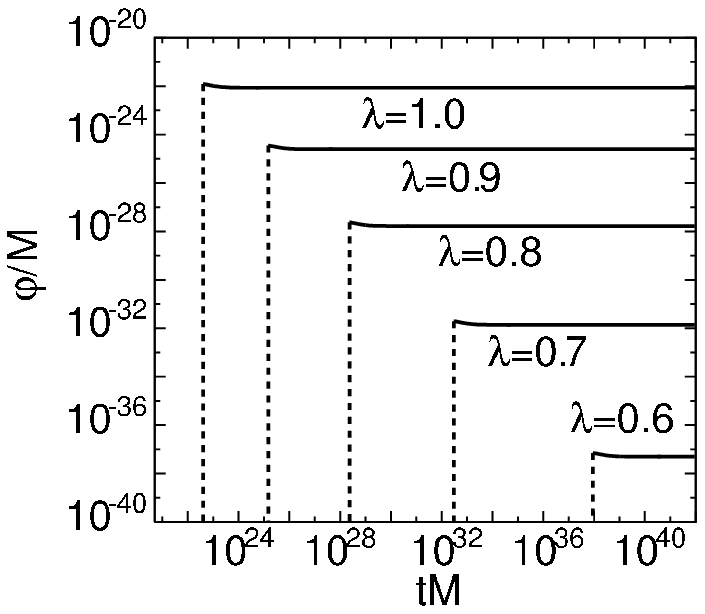}
(c) $\alpha$=0, $\xi=0$
\end{center}
\end{minipage}
\begin{minipage}{6.8cm}
\begin{center}
\includegraphics[width=\linewidth]{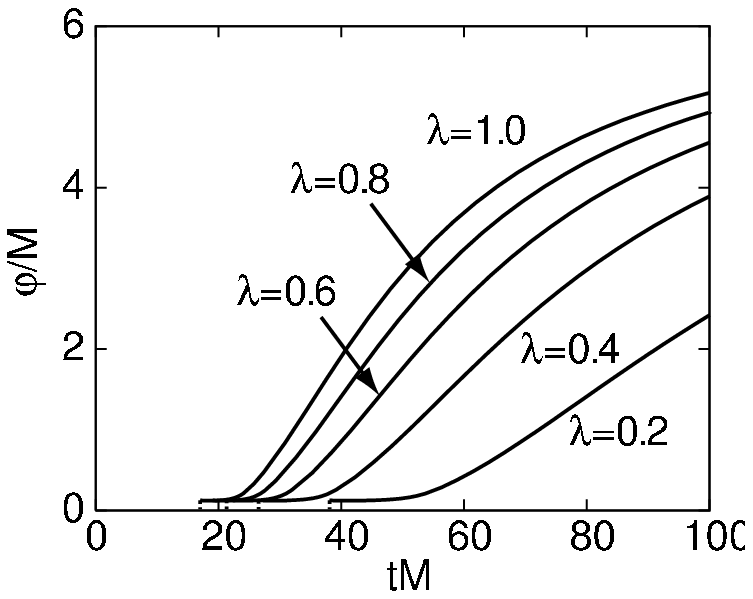}
(d) $\alpha$=1, $\xi=0$
\end{center}
\end{minipage}
\caption{\label{gapminimu0c2}Behaviour of the mass
gap for $h_0=2$, $\mu_r=0$ and $D=4$.}
\end{center}
\end{figure}

\begin{figure}[t]
\begin{center}
\begin{minipage}{6.8cm}
\begin{center}
\includegraphics[width=\linewidth]{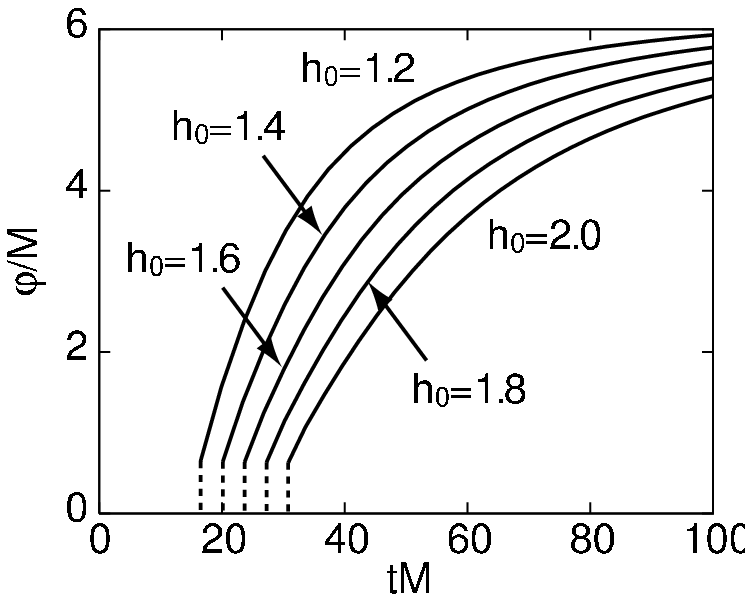}
(a) $\xi=\xi_{\mbox{\scriptsize conformal}}$
\end{center}
\end{minipage}
\begin{minipage}{6.8cm}
\begin{center}
\includegraphics[width=\linewidth]{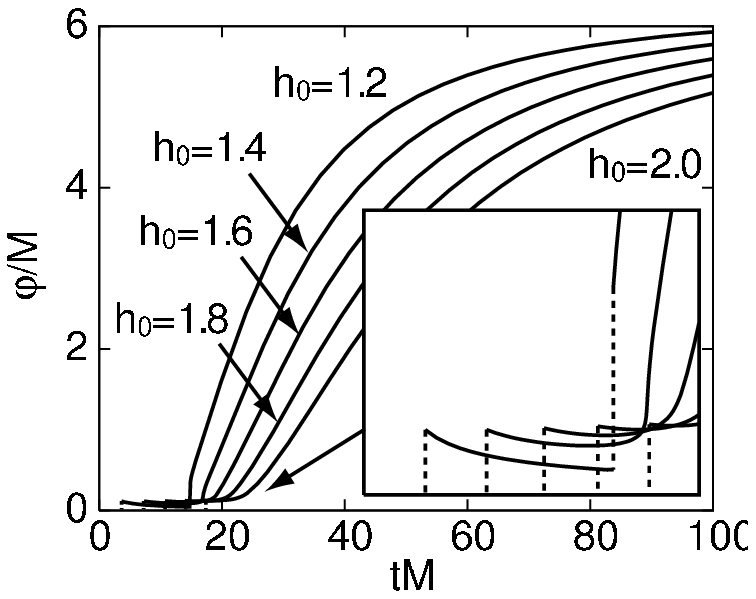}
(b) $\xi=0$
\end{center}
\end{minipage}
\caption{\label{gapmu0a1l1}Behaviour of the mass
gap for $\alpha=1$, $\mu_r=0$, $\lambda=1$ and $D=4$.}
\end{center}
\end{figure}

\begin{figure}[t]
\begin{center}
\begin{minipage}{6.8cm}
\begin{center}
\includegraphics[width=\linewidth]{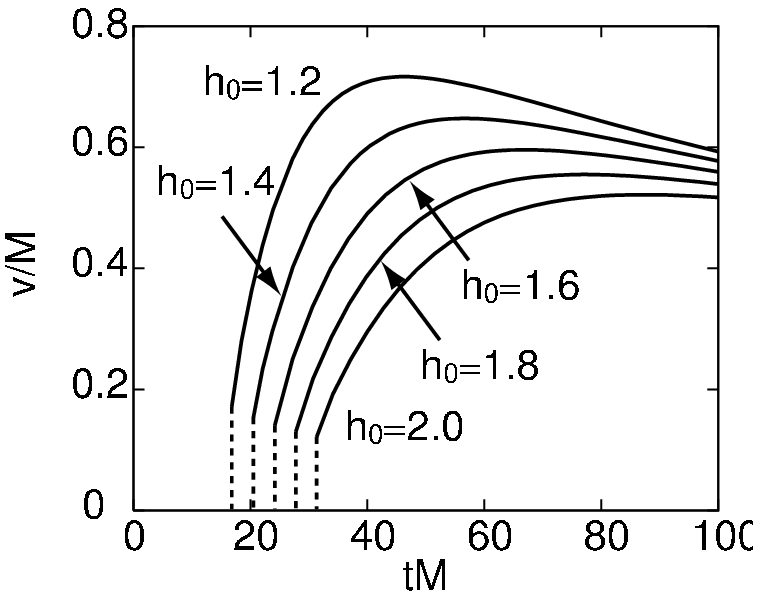}
(a) $\langle \phi \rangle = v \sqrt{t}$, 
$\xi=\xi_{\mbox{\scriptsize conformal}}$
\end{center}
\end{minipage}
\begin{minipage}{6.8cm}
\begin{center}
\includegraphics[width=\linewidth]{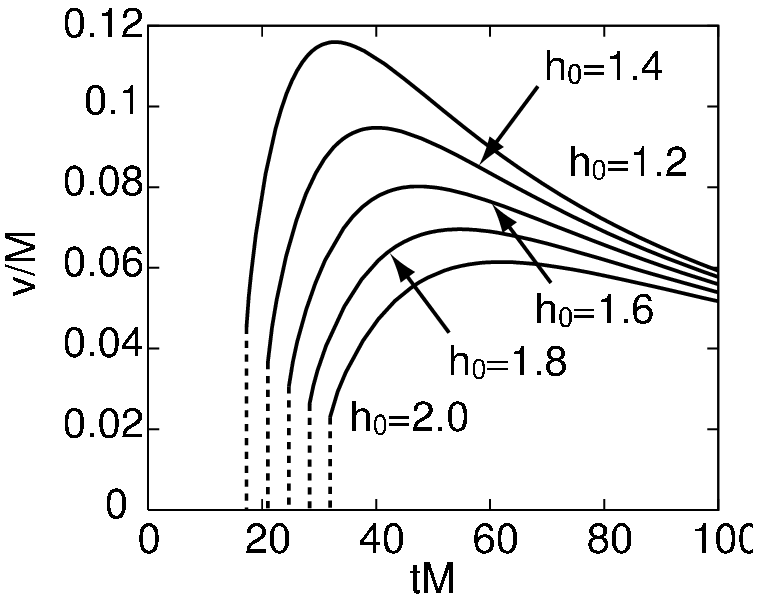}
(b) $\langle \phi \rangle = v t$, 
$\xi=\xi_{\mbox{\scriptsize conformal}}$
\end{center}
\end{minipage}
\begin{minipage}{6.8cm}
\begin{center}
\includegraphics[width=\linewidth]{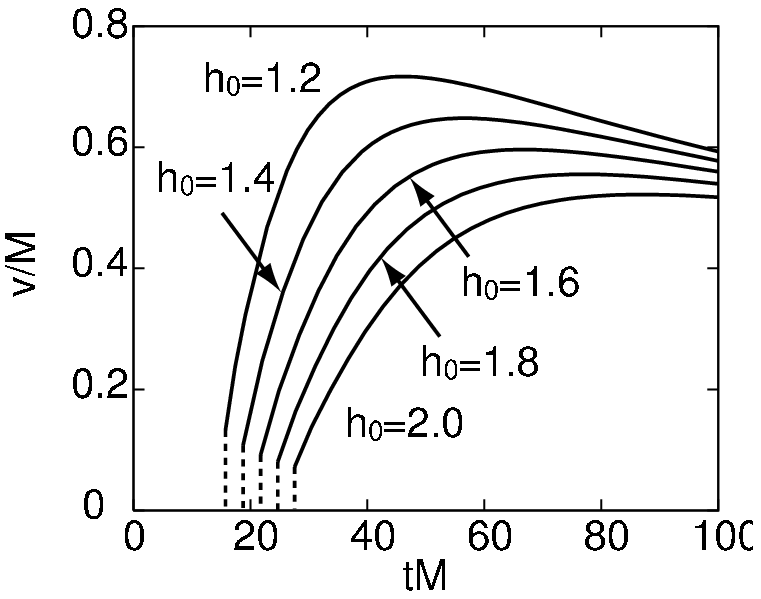}
(c) $\langle \phi \rangle = v \sqrt{t}$, $\xi=0$ 
\end{center}
\end{minipage}
\begin{minipage}{6.8cm}
\begin{center}
\includegraphics[width=\linewidth]{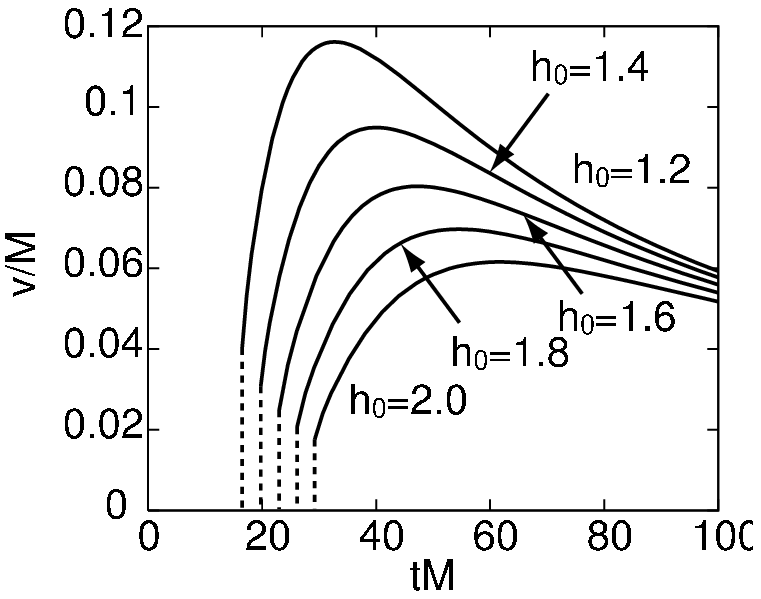}
(d) $\langle \phi \rangle = v t$, $\xi=0$ 
\end{center}
\end{minipage}
\caption{\label{minimu0a1c1d1}Behaviour of the mass scale $v$ for $\alpha=1$,
$\mu_r=0$, $\lambda=1$ and $D=4$.}
\end{center}
\end{figure}

In Fig. 1 we illustrate the behavior of the effective potential for a 
conformal and a minimal gravitational coupling $\xi=1/6$ and $\xi=0$ 
respectively.
It should be noted that the theory is conformally invariant only when 
$\alpha=0$.
Behaviors of the minimal of the effective potential are drawn
in Figs. 2 and 3 with varying $\lambda$ and $h_0$.
First order phase transition is observed in Figs. 2 and 3.
However, the radiative correction has only a small effect in an 
ordinary theory. The expectation value generated by the radiative symmetry 
breaking is extremely small at $\alpha=0$, as is shown in Fig. 2. 
For a positive $\alpha$ the non-linear curvature coupling enhances radiative 
corrections as curvature decreases. Thus the expectation value 
$\langle\phi\rangle$ becomes extremely larger in comparison with that 
for the case $\alpha=0$.
In Fig. 3 we observe two steps of the transition at $h_0=1.2$.
In the case of negative $\alpha$ the radiative correction is suppressed 
as the curvature decreases and then radiative symmetry breaking does not 
occur.

As is shown in Figs. 2 and 3, the expectation value $\langle\phi\rangle$ has
non-negligible time dependence, especially for $\alpha=1$.
Hence we next consider the spatially homogeneous but time dependent $\phi$ at
$\alpha =1$.
The time evolution of $\langle\phi\rangle$ is described by
the equation of motion:
\begin{eqnarray}
    && \frac{1}{\sqrt{-g}}
    \frac{\delta S}{\delta \phi} = -\frac{\partial V(\phi)}{\partial \phi}
        -\left(\frac{R}{M^2} \right)^{\alpha} \left[
         \ddot{\phi}+3\frac{\dot{a}}{a}\dot{\phi}
         -\frac{2\alpha}{t}\dot{\phi}
        \right]
     =  0 .
\label{eqm}
\end{eqnarray}
To find an exact solution one needs to solve this equation for a general
time-dependent form of $\phi$. However, it is instructive to consider the
solution of Eq.(\ref{eqm}) for a special form of $\phi$.
In the present paper it is assumed that
\begin{equation}
      \phi(t)=\langle\phi(t)\rangle = v t^x,
      \ \ \phi^{;\mu}\phi_{;\mu} = \frac{x^2}{t^2} \langle \phi(t) \rangle^2 ,
\label{ass2}
\end{equation}
where $v$ is a constant parameter. Here we fix the parameter
$x$ at $1/2$ or $1$ and numerically solve the equation of motion (\ref{eqm}).

The solution is shown in Fig. 4.
As is seen in Fig. 4, we observe the first order phase transition again.
In this case the mass scale $v$ depends on the time t again. However, we 
observe that the mass scale $v$ is almost static around $tM\sim 60$ for 
$x=1$ and larger $t$ for $x=1/2$. It is expected that there is a solution 
with gradually decreasing $x$ after the first order transition. 
It is also interesting that the $\xi$ dependence of $v$ is smaller than
results for a stationary $\phi$.

\section{Resolution of the cosmological constant problem}

There are some proposal for solving the cosmological constant problem 
dynamically \cite{dolgov, J}.
One of the possible solutions was pointed out by Mukouyama and Randall 
\cite{MR}. 
Here we consider the scalar theory non-linearly coupled with the curvature
(\ref{a0}) in the four dimensional FRW metric with flat spatial part. 
We apply the analysis by Mukouyama and Randall to our model. 
A solvable case is found at $\alpha=-1$ \cite{INO}. 

As is shown in the previous section, the radiative correction is suppressed 
when the curvature is small for a negative $\alpha$. The behavior 
of the scalar field $\phi$ is found by solving the Einstein equation and the 
field equation for $\phi$ simultaneously,
A solution of these equations is given by
\begin{equation}
\label{MRIO19}
H=\frac{h_0}{t},\ \phi=\frac{\phi_0}{t}, \ (h_0>0),\ \  \mbox{ or } \ \ 
H=\frac{h_0}{t_s-t},\ \phi=\frac{\phi_0}{t_s-t}, \ (h_0<0).
\end{equation}
Substituting (\ref{MRIO19}) into the Einstein equation and the field 
equation, we find the solution:
\begin{eqnarray}
\label{MRIO22}
\phi_0^2&=&\frac{3}{\displaystyle\kappa^2\left\{\frac{8-9h_0}{24\left(-h_0 +
h_0^2\right)^2}
- \frac{\left(4-7h_0\right)\xi}{\left( -h_0 + 2h_0^2\right)h_0}\right\}}\ ,\\
\label{MRIO23}
\lambda&=& -6\kappa^2h_0 \left\{ 1 - 2\left(1-2h_0\right)\xi\right\} \nonumber
\\
&& \times \left\{\frac{8-9h_0}{24\left(-h_0 + h_0^2\right)^2}
- \frac{\left(4-7h_0\right)\xi}{\left( -h_0 + 2h_0^2\right)h_0}\right\}\ .
\end{eqnarray}

For example we consider a minimal coupling case, $\xi=0$, here.  
Since $\phi_0^2$ should be positive, one finds
$h_0 \leq 9/8$.

If we choose $h_0=-1/60$, it gives the state equation parameter $w$ as
\begin{equation}
 w = -1+\frac{2}{3 h_0}=-1.025,
\end{equation}
It is consistent with the observed one.
Therefore, with the proper choice of parameters we obtain the 
solution for the Einstein equation and the field equation.

As is clearly seen the Habble rate $H$ is suppressed as time runs. 
If we substitute the present age of the 
Universe $10^{10}$year into $t$ or $t_s-t$, the observed value of $H$ could 
be reproduced. It explains the smallness of the effective cosmological 
constant $\Lambda\sim H^2$. 
Then by properly choosing the parameters, we may obtain an exact solution for
cosmological constant.

\section{Concluding remarks}

We have investigated the radiative symmetry breaking and discuss the dynamical 
resolution of the cosmological constant problem in the scalar 
self-interacting theory non-linearly coupled with some power of the 
curvature. We numerically evaluate the one-loop effective Lagrangian in the 
four dimensional FRW spacetime with flat spatial part at the weak curvature
limit. The phase structure strongly depends on the sign of $\alpha$. 
The $\xi$ dependence of it is very weak. 
For a non-negative 
$\alpha\ (\geq 0)$ 
we observed the first-order phase transition. Compared with the usual 
$\phi^4$ theory in weakly curved space, i.e. $\alpha=0$, the 
expectation value $\langle\phi\rangle$ is extremely enhanced for a positive
$\alpha$. In the case of a negative $\alpha$ the radiative correction is 
suppressed as curvature decreases. No radiative symmetry breaking 
is observed for $\alpha<0$. 
We apply the mechanism proposed in Ref.\cite{MR} to our model with 
negative $\alpha$.
Dynamical mechanism to solve the cosmological constant problem 
is naturally realized also for the class of models investigated 
in Ref. \cite{INO}.

As the simplest model we consider radiative symmetry breaking of a 
discrete $Z_2$ symmetry. Breaking of the discrete symmetry must construct
a domain wall structure in our universe. It is only a prototype model
of the dark energy. We do not expect to explain all the problem in 
this simplest model. It is straightforward to perform the same analysis 
in a complex scalar theory which has a continuous $U(1)$ symmetry. 
Then we can avoid the domain wall problem and find the same behaviors.

There are many directions where our approach may be generalized. 
In particularly, it is known that phase structure in the NJL-like model 
in curved spacetime is quite rich (for a review, see \cite{IMO}). 
It is quite interesting to consider models with fermions and gauge bosons.
From another point of view, the new matter-gravity coupling \cite{NO} may 
also be considered as a kind of modification of gravitation itself. 
It should be interesting to calculate the one-loop effective action when 
the extreme gravity is formulated in Palatini form. That attracts us to 
further research works continuously.

\ack
The main part of this paper is based on the work \cite{INO}.
The author benefited a lot from discussions with S. Nojiri and 
S. D. Odintsov.


\section*{References}

\begin{thebibliography}{99}
\bibitem{INO}
    Inagaki T, Nojiri S and Odintsov S D 2005 JCAP 0506(2005)010.
\bibitem{NO}
    Nojiri S and Odintsov S D 2004 \PL B {\bf 599} 137
\bibitem{ANO}
    Abdalla M C B, Nojiri S and Odintsov S D 2005
    \CQG {\bf 22} L35
\bibitem{BOS}
    Buchbinder I L, Odintsov S D and Shapiro I L 1992
    {\it Effective Action in Quantum Gravity} (IOP Publishing)
\bibitem{P}
    Petrov A Z 1969
    {\it Einstein Space}, (Pergamon, Oxford)
\bibitem{BP}
    Bunch T S and Parker L 1979 \PR D {\bf 20} 2499
\bibitem{PT}
    Parker L and Toms D J 1984 \PR D {\bf 29} 1584
\bibitem{dolgov}
    Dolgov A D and Kawasaki M 2003 {\it Preprint} astro-ph/0307442
\bibitem{J}
    Jackiw R, Nunez C and Pi S -Y 2005 \PL A {\bf 347} 47 
\bibitem{MR}
    Mukohyama S and Randall L 2004 \PRL {\bf 92} 211302
\bibitem{IMO}
   Inagaki T, Muta T and Odintsov S D 1997
   {\it Prog. Theor. Phys. Suppl.} {\bf 127} 93
\end{thebibliography}
\end{document}